\documentclass[%
 reprint,
 amsmath,amssymb,
 aps,
]{revtex4-2}

\usepackage{graphicx}
\usepackage{dcolumn}
\usepackage{bm}
\usepackage{hyperref}
\usepackage{float}
\usepackage[svgnames]{xcolor}
\usepackage{comment}
\newcommand{\orcid}[1]{\href{https://orcid.org/#1}{\includegraphics[width=8pt]{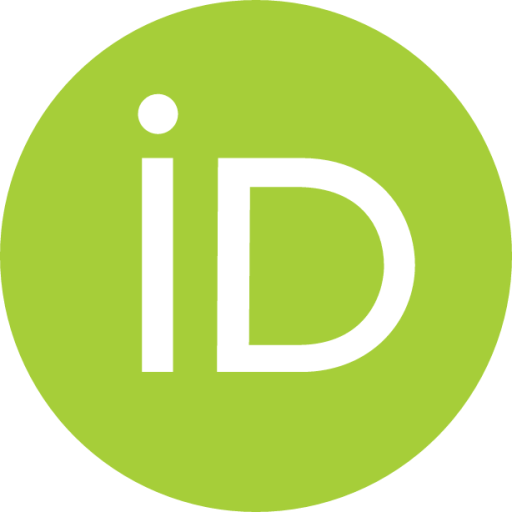}}}

\newcommand{\myst}[1]{%
    \scriptscriptstyle{\text{#1}}
    }

\begin{document}

\title{Enhancement of Electromagnetic Memory Effects}

\author{Jann Zosso\orcid{0000-0002-2671-7531}}
\email{jann.zosso@nbi.ku.dk}
\affiliation{Center of Gravity, Niels Bohr Institute, Blegdamsvej 17, 2100 Copenhagen, Denmark}
\affiliation{Albert Einstein Center, Institute for Theoretical Physics, University of Bern, Sidlerstrasse 5, 3012, Bern, Switzerland}

\date{\today}

\begin{abstract}
We show that the amplitude of electromagnetic memory can be significantly enhanced in comparison to known estimates. In a Lorentz breaking phase of lowered phase velocity of light, there exist critical spacetime directions of memory-source emission along the effective light cone, about which the total memory offset receives order of magnitude increases. The same amplification is already present in the Lorentz preserving case by considering ultrarelativistic memory-source charges. These observations may pave the way for a first observation of the phenomenon of memory and laboratory tests of the concepts of asymptotic symmetries and soft theorems.
\end{abstract}

\maketitle


\section{\label{sec:Intro}Introduction}






Gravitational radiation not only consists of the observed gravitational waves, but always also includes a nonwavelike component that permanently alters the background structure of asymptotically flat spacetime in the form of a net shift in proper distances. This phenomenon is known as \emph{gravitational memory} \cite{Zeldovich:1974gvh,Braginsky:1985vlg, Braginsky:1987gvh,Christodoulou:1991cr,Blanchet:1992br,FrauendienerJ,Thorne:1992sdb,PhysRevD.44.R2945} (see also \cite{Favata:2009ii,Favata:2010zu,Strominger:2014pwa,Bieri:2015yia,Pasterski:2015tva,Garfinkle:2022dnm,Zosso:2025ffy}) and represents one of the fundamental predictions of general relativity that has not yet been observed. This additional contribution to the radiation can be viewed as a back-reaction effect of emitted radiative energy momentum. The resulting memory-induced transition on the state of the background spacetime is radically related to a supertranslation of asymptotic Bondi-Metzner-Sachs (BMS) symmetries \cite{Bondi:1962px,Sachs:1962wk,FrauendienerJ,Ashtekar:2014zsa,Strominger:2014pwa,Compere:2019gft} and can therefore be viewed as a spontaneous breaking of the latter. Moreover, as a ``net'' background effect, memory probes the zero frequency limit of the radiating gauge field, such that the phenomenon is equally connected to the gravitational soft theorems \cite{Weinberg_PhysRev.140.B516,Strominger:2014pwa,Strominger:2017zoo}.

The relation to soft factors, as well as the fact that gravitational memory can be viewed as a mere consequence of the wave equation in a radiative process \cite{PhysRevD.44.R2945,Garfinkle:2022dnm,Heisenberg:2023prj,Zosso:2024xgy,Zosso:2025ffy}, suggests that an analogous phenomenon must exist within the theory of electromagnetism. Indeed, based on the gravitational knowledge, it was possible to describe an \emph{electromagnetic memory} \cite{Bieri:2013hqa} that manifests itself as a nonzero offset within the radiative transverse vector potential, representing the classical manifestation of photon soft theorems and the asymptotic $U(1)$ symmetry of electric charge conservation \cite{He:2014cra,Campiglia:2015qka,Pasterski:2015zua,Kapec:2015ena,Strominger:2017zoo}. In contrast to the gravitational case, however, electromagnetic waves themselves do not carry any corresponding charge, such that a production of electromagnetic memory necessarily requires an emission of unbound charged matter.

While the first single-event detection of gravitational memory is only expected with next generation gravitational wave detectors \cite{Grant:2022bla,Inchauspe:2024ibs}, the electromagnetic analog provides the possibility to probe this fundamental aspect of radiation theory, together with the deep implications on asymptotic symmetries and soft theorems, in a controllable laboratory setting \cite{Susskind:2015hpa,Bieri:2023btq}. However, no concrete efforts to observe electromagnetic memory have been undertaken yet. 

In gravity, it was recently shown that a local Lorentz breaking of the dynamical field equations results in a mechanism for producing significantly increased memory amplitudes along specific critical spacetime directions \cite{Heisenberg:2025tfh}.
In this work, we perform the computation of memory for superluminal electric charge emission in section~\ref{sec:MemoryCalc} and demonstrate that the same mechanism of memory enhancement is present in the electromagnetic theory. In contrast to the gravitational case, a significant increase in amplitude is even already present in Lorentz preserving vacuum as the velocity of charge emission asymptotes towards the speed of light $c$. First observational consequences are discussed in section~\ref{sec:ObservationalImplications} with concluding thoughts in section~\ref{sec:Conclusion}. Appendix~\ref{App:FullMemorySolution} gathers the full expressions of the solutions and deals with the freedom in the polarization angle. Appendix~\ref{sSec:IndirectObservation} additionally computes the energy spectrum of the radiated memory signal and offers a connection a known model for the observed radiation in beta decay, while in appendix~\ref{App:LightconeEmission} we provide a geometric understanding of the critical spacetime directions of memory enhancement. In order to track factors of $c$, we will work in SI units.


\section{\label{sec:MemoryCalc}Memory from a Lorentz Breaking Wave Equation}

Although the main conclusions in this section hold for any form of reduction in the phase velocity of light $v_\text{em}<c$, for the sake of concreteness, we will present the computation in the context of a simple dielectric medium. Furthermore, we will focus on the emission of a single charge $q$ from a localized source, which provides the basic building block of the memory formula, from which multiple charge scenarios can be deduced in an additive manner.

\subsection{Dynamical Memory Equation}

The dynamical equation of the electromagnetic vector potential in a Lorentz breaking simple dielectric medium with free source $J^\alpha$, reads in Minkowski coordinates $\{t,x,y,z\}$ and in Lorenz gauge
\begin{equation}\label{eq:MaxwellDielectric}
    -\frac{1}{v_\text{em}^2}\ddot A^\alpha+\,\partial^2 A^\alpha =-\mu J^\alpha\,.
\end{equation}
The effective phase velocity of light 
\begin{equation}
    v_\text{em}\equiv \frac{c}{n}\,,
\end{equation}
is characterized by the refractive index of the medium
\begin{equation}
    n\equiv \frac{\sqrt{\epsilon\mu}}{\sqrt{\epsilon_0\mu_0}}\,,
\end{equation}
that depends both on the permittivity $\epsilon$ and permeability $\mu$. The mere fact that the vector potential satisfies a flat spacetime wave equation implies the existence of an associated electromagnetic memory effect \cite{Garfinkle:2022dnm}. However, the breaking of Lorentz symmetry by a dielectric medium considerably alters the implications, as we will now show.

Let's assume the idealized model of a fast electric charge $q$ that is instantly emitted from a localized process at the coordinate origin with constant 3-velocity $\frac{d\mathbf{x}_q}{dt}=v_q \hat{\mathbf n}_q$ in a given direction characterized by a unit radial outward vector $\hat{\mathbf n}_q=\hat{\mathbf n}(\Omega_q)$ defined through the angles $\Omega_q=\{\theta_q,\phi_q\}$ in spherical coordinates [Eq.~\eqref{eq:Def Direction of Propagation n}]. The associated 4-velocity reads $U_q^\alpha=\frac{d x_q^\alpha(\tau)}{d\tau}=\gamma_q V^\alpha_q $, where $V^\alpha_q\equiv(v_\text{em},v_q\hat{\mathbf n}_q)$ and $\gamma_q\equiv 1/\sqrt{1-(v_q/c)^2}$ and the 4-current density becomes
\begin{equation}\label{eq:SourceCurrent}
    J_q^\alpha(t,\vec{x})=q\int d\tau\, U_q^\alpha\,\delta^4(x-x_q(\tau))=q\,V_q^\alpha\delta^3(\mathbf{x}-\mathbf{x}_q(t))\,.
\end{equation}
The emission of such an unbounded charge inevitably sources a component in the asymptotic radiation that can be computed by solving Eq.~\eqref{eq:MaxwellDielectric} within the asymptotic limit.

\subsection{A General Electromagnetic Memory Solution}

The solution for the corresponding radiative transverse components $A_T^i$ is obtained through 
\begin{equation}
    A_T^i=-\mu\perp^{ij}\int d^4x'G(x-x')J_q^j(x')\,,
\end{equation}
where the transverse projector is defined as
\begin{equation}
    \perp^{ij}\equiv \delta^{ij}-\hat n^i \hat n^j\,,
\end{equation}
and the appropriate retarded Green's function reads
\begin{equation}
    G(x-x')=-\frac{\delta(t_{\text{ret}}-t')}{4\pi\,|\vec{x}-\vec{x}'|}\,,
\end{equation}
with retarded time defined as
\begin{equation}
    t_{\text{ret}}\equiv t-\frac{|\vec{x}-\vec{x}'|}{v_\text{em}}\,.
\end{equation}
To evaluate the expression in the limit to null infinity, it is convenient to switch to spherical coordinates 
\begin{equation}
    \{t,x,y,z\}\rightarrow  \{u,r,\theta,\phi\}\,,\; \{t',x',y',z'\}\rightarrow  \{u',r',\theta',\phi'\}\,,
\end{equation}
with the time variable replaced by the asymptotic retarded times
\begin{equation}
    u\equiv t-\frac{r}{v_\text{em}}\,,\qquad
    u'\equiv t'-\frac{r'}{v_q}\,.
\end{equation}
With the assumption $r'\ll r$ the Green's function becomes to leading
order in $r'/r$
\begin{equation}\label{eq:GreensfunctionFinal}
    G(x-x')=
    -\frac{v_q|\mathcal{V}|}{4\pi r}\,
    \delta\!\left(r'-(u-u')v_q\mathcal{V}\right)\,,
\end{equation}
where we define the dimensionless ratio
\begin{equation}\label{eq: Greensfunction factor}
    \mathcal{V}\equiv
    \frac{1}{
    1-\frac{v_q}{v_\text{em}}\,
    \hat{\mathbf n}\cdot\hat{\mathbf n}'}\,.
\end{equation}

On the other hand, parameterizing the radial outward trajectory of the source charge as
\begin{equation}
    \mathbf{x}_q(t')=v_q\,\hat{\mathbf n}_q\,(t'-u_q)\,,
\end{equation}
defining worldlines of constant source retarded time $u_q$, the source current in Eq.~\eqref{eq:SourceCurrent} can be written as
\begin{equation}\label{eq:ChargeCurrent}
    J^i(u',r',\Omega')=\frac{q\, \hat n^i}{r'^2}\delta(u'-u_q)\delta^2(\Omega'-\Omega_q)\,.
\end{equation}
Depending on the sign of the Green's function factor $\mathcal{V}$ in Eq.~\eqref{eq: Greensfunction factor}, one obtains two different solutions. 

\textbf{Case (I). $\mathcal{V}>0$ :}
In this case, the delta function in Eq.~\eqref{eq:GreensfunctionFinal} dictates $u'\leq u$ and the final result reads
\begin{equation}\label{eq:SolMemI}
    _{\myst{(I)}}A_T^i(u,r,\Omega)=\frac{q}{4\pi r} \mu\, v_q \Theta(u-u_q) \frac{\perp^{ij}\hat n_q^j}{1-\frac{v_q}{v_\text{em}}\hat{\mathbf{n}}\cdot\hat{\mathbf n}_q}\,,
\end{equation}
with $\Theta$ the Heaviside stepfunction. Observe that this component in the asymptotic radiation indeed represents an electromagnetic memory within the transverse vector potential. In the vacuum, $v_\text{em}=c$, this result coincides with the electromagnetic memory computed in \cite{Bieri:2013hqa,Garfinkle:2022dnm}.

\textbf{Case (II). $\mathcal{V}<0$ :} In a dielectric medium with $v_\text{em}<v_q\leq c$, there exists a parameter space, in which the Green's function factor $\mathcal{V}$ changes sign. In this case, the delta function in Eq.~\eqref{eq:GreensfunctionFinal} implies $u'\geq u$ \footnote{See \cite{Heisenberg:2025tfh} for a causal interpretation of this statement.} which translates into a memory formula of 
\begin{equation}\label{eq:SolMemII}
    {}_{\myst{(II)}}A_T^i(u,r,\Omega)=
    \frac{q\,\mu v_q}{4\pi r}
    [\Theta(u-u_q)-1]
    \frac{\perp^{ij}\hat n_q^j}
    {1-\frac{v_q}{v_\text{em}}
    \hat{\mathbf n}\cdot\hat{\mathbf n}_q}\,.
\end{equation}
Hence, if the velocity of the emitted charge is greater than the effective velocity of light within the dielectric medium, there exist two parameter space regions with different solutions.

\subsection{Directions of Memory Enhancement}\label{sSec:DirecitonsOfMemoryEnhancement}
The two case regions (I) and (II) introduced above are separated by a diverging line $|\mathcal{V}|\rightarrow \infty$ characterized by the condition
\begin{equation}\label{eq:LightconeCond}
    \hat{\mathbf{n}}\cdot\hat{\mathbf{n}}_q=\frac{v_\text{em}}{v_q}\,.
\end{equation}
This divergence is an indication for a large physical enhancement of the memory amplitude around the transition region \cite{Heisenberg:2025tfh}. As shown in Fig.~\ref{fig:DensityPlotEMem}, the magnitude for the total memory amplitude as computed in appendix~\ref{App:FullMemorySolution} [Eq.~\eqref{eq:TotalMemOffset}] in a the superluminal phase
\begin{equation}\label{eq:defememsup}
    \Delta A^T_\text{sup}\equiv \Delta A^T|_{v_q=0.99 c,\,v_\text{em}=0.9 c}\,,
\end{equation}
is amplified by several order of magnitude around the diverging line compared to a subluminal solution
\begin{equation}\label{eq:defememsub}
    \Delta A^T_\text{sub}\equiv \Delta A^T|_{v_q=0.88 c,\,v_\text{em}=0.9 c}\,.
\end{equation}
 Note that the order of magnitude of memory amplification does not depend on the particular values of velocities. Only the angle of the critical direction depends through Eq.~\eqref{eq:LightconeCond} on the ratio of charge and phase velocities.
\begin{figure}[H]
  \centering
    \includegraphics[scale=0.36]{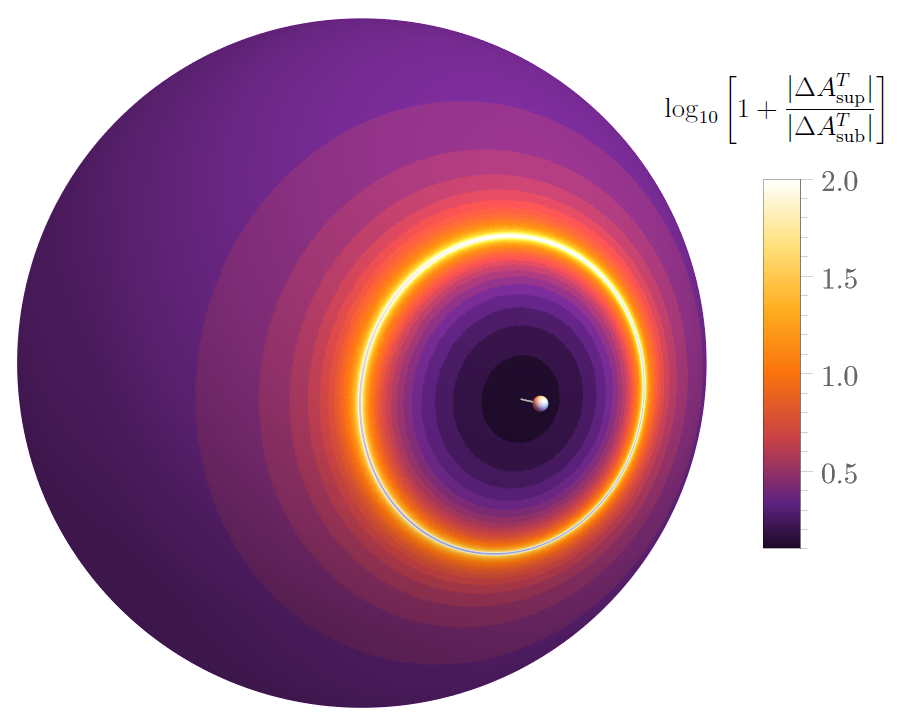}
  \caption{\small The ratio of the magnitudes of the total memory offsets
$|\Delta A^T_\text{sup}|/|\Delta A^T_\text{sub}|$ with superluminal speed $v_q=1.1 \,v_\text{em}$ [Eq.~\eqref{eq:defememsup}] and with subluminal value $v_q=0.978\,v_\text{em}$ [Eq.~\eqref{eq:defememsub}] as a density plot over the sphere of memory emission centered at the position of charge emission. The spacial direction of charge emission is indicated by a thin white line with the superluminal charge represented by a white particle. For a superluminal emission, there exists critical angles $\hat{\mathbf{n}}\cdot\hat{\mathbf{n}}_q=v_\text{em}/v_q$, around which the amplitude of electromagnetic memory is enhanced by several orders of magnitude. For visibility, the plot is clipped at the scale of an increase in two orders of magnitude.}\label{fig:DensityPlotEMem}
\end{figure}
\noindent

\begin{figure}[H]
  \centering
    \includegraphics[scale=0.69]{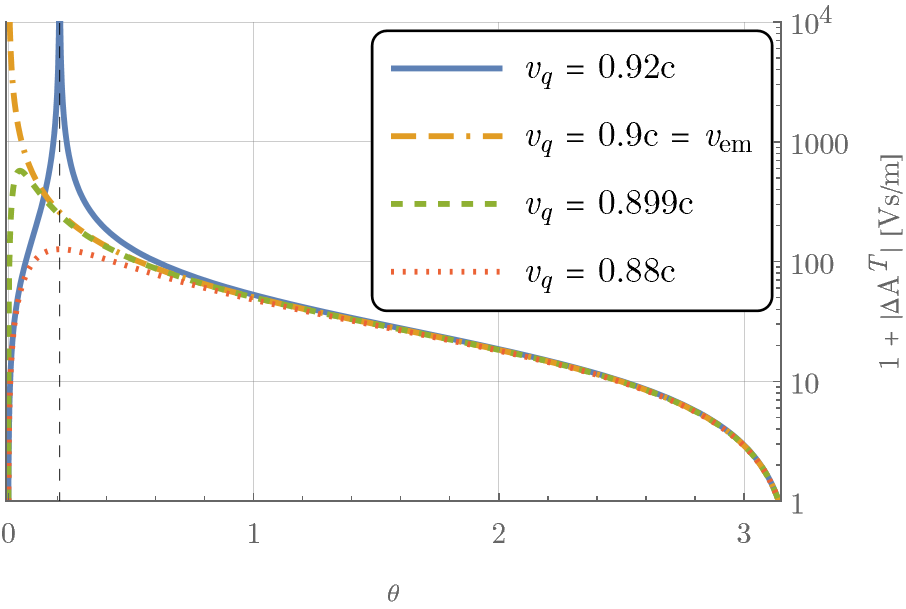}
  \caption{\small The magnitude of the total electromagnetic memory offset
$|\Delta A^T(1,\theta,\theta_q=0)|$ [Eq.~\eqref{eq:TotalMemOffsetSimp}] in SI units of [Vs/m] for a unit charge $q=1$[C] and distance $r=1$[m], as a function of the polar angle $\theta$ and for different values of charge emission speeds $v_q$. For concreteness, the phase velocity of light in the medium is set to $v_\text{em}=0.9c$, with $\mu=\mu_0$ and $\epsilon\approx 1.23 \,\epsilon_0$. For a superluminal emission of the charge $v_q>v_\text{em}$ (solid blue) there exists an unprotected direction of unbounded memory sourcing $\theta_c=\arccos[v_\text{em}/v_q]=0.2089$, indicated by a black dashed vertical line, around which the amplitude of electromagnetic memory is considerably enhanced. If $v_q=v_\text{em}$ (dotdashed orange) the critical angle coincides with the direction of emission of the source $\hat{\mathbf{n}}=\hat{\mathbf{n}}_q$. For a subluminal emission velocity $v_q<v_\text{em}$ (dashed green and dotted red), the electromagnetic memory offset remains finite for any values of $\theta$, although the amplitude is still significantly increased at low values of the angle $\theta$ as soon as $v_q \gtrsim 0.99\, v_\text{em}$.}\label{fig:PlotEMem}
\end{figure}

Since the solution is $U(1)$ symmetric about the axis of emission of the charged particle, it is useful to choose an emission along the $z$-direction of the given coordinate system by setting $\theta_q=0$, in which case the total amplitude of electromagnetic memory is entirely parameterized by the polar angle $\theta$ and $\hat{\mathbf{n}}\cdot\hat{\mathbf{n}}_q=\cos\theta$. The resulting magnitude of the memory offset within the vector potential [Eq.~\eqref{Appeq:TotalMemOffsetSimp}] 
\begin{align}\label{eq:TotalMemOffsetSimp}
    \Delta A^T(r,\theta,\theta_q=0)
    &=
    \frac{q\mu v_q}{4\pi r}
    \frac{\sin\theta}
    {1-\frac{v_q}{v_\text{em}}\cos\theta}\,,
    \;\,
    \frac{v_q}{v_\text{em}}\cos\theta\neq1\,.
\end{align}
is visually represented in Fig.~\ref{fig:PlotEMem} for different values of the charge propagation speed $v_q$. For concreteness, we have chosen a reduced speed of light in the medium of $v_\text{em}=0.9c$, with $\mu=\mu_0$ and $\epsilon\approx 1.23 \,\epsilon_0$.

Observe that, in contrast to the gravitational case \cite{Heisenberg:2025tfh}, the memory amplitude enhancement is already present in the subluminal regime as the speed of the emitted charge $v_q$ approaches the phase velocity of light $v_\text{em}$. This is because the rate at which the memory amplitude asymptotes to zero in the direction of charge emission, as dictated by transverse projection, is slower in the electromagnetic case. More precisely, one can estimate that the distinctive feature close to the critical direction starts appearing for $v_q\gtrsim 0.99\, v_\text{em}$. A significant increase in memory amplitude can therefore also be achieved with vacuum speed of light $v_\text{em}=c$ for ultra-relativistic memory source charges. However, only in Lorentz breaking superluminal configurations does the direction of enhanced emission break free from the particular direction of charge emission, which might be crucial for concrete experimental setups.

While the divergence in the memory amplitude at the exact critical direction in Eq.~\eqref{eq:LightconeCond} is unphysical - it directly translates into a divergence in the emitted radiative energy as shown in appendix~\ref{sSec:IndirectObservation} and the underlying assumptions of memory computations break down - the significant increase in memory offset in the vicinity of the critical directions is real \cite{Heisenberg:2025tfh}. This is also shown by the fact that the enhancement already occurs in an entirely finite and well-defined manner within a Lorentz-preserving vacuum phase, as discussed above.
Furthermore, we explicitly show in appendix~\ref{App:LightconeEmission} that the divergence condition can geometrically be explained as arising whenever the source charge is asymptotically emitted along the spacetime direction of a past null cone characterized by the phase velocity of light $v_\text{em}$. Such critical directions will therefore only be present at luminal or superluminal speeds $v_q$ of charge emission. We note that Eq.~\eqref{eq:LightconeCond} is equivalent to the condition on the emission angle of Cherenkov radiation \cite{Cherenkov:1934ilx,V_Jelley_1955,Ratcliff:2021ofp} \footnote{The crucial difference to Cherenkov radiation is that the actual production of memory radiation only occurs during the acceleration process of the source}. The above arguments strongly suggest that the enhanced memory burst is a shock-wave phenomenon, generally characterized by the presence of a formally singular potential.


\section{\label{sec:ObservationalImplications}Observational Implications}

In analog to the geodesic deviation equation in gravity, it is the Lorentz force on a test charge $Q$ with mass $M$
\begin{equation}
    M\frac{d^2\mathbf{x}}{dt^2}=Q\mathbf{E}\,,
\end{equation}
that reveals the observational implications of a nonzero offset within the radiative vector potential $\Delta A_T^i\neq 0$.
Assuming a finite burst of electromagnetic radiation, an integration of this equation results in
\begin{equation}
    \Delta v^i=\frac{Q}{M}\int_{-\infty}^\infty dt E^i=-\frac{Q}{M}\Delta A_T^i\,,
\end{equation}
where we have used
\begin{equation}\label{eq:ElectricFieldtoA}
    E_T^i=-\partial A_T^i/\partial t\,.
\end{equation}
Hence, once all electromagnetic radiation has passed, the test charge $Q$ receives a nonzero transverse velocity kick \cite{Bieri:2013hqa}\footnote{To prevent confusion we want to briefly mention here that the concern about the non-existence of the velocity kick on asymptotic test charges expressed in Ref.~\cite{Herdegen:2023qdw} is based on a misunderstanding: While that reference computes the Coulombic field of outgoing free charged particles, and therefore correctly infers that in such a physical situation there is no energy-carrying radiation, the electromagnetic memory effect is a locally produced radiation field that requires the acceleration or scattering of unbound charges.}. This memory kick is distinguished from radiation pressure associated to the Poynting vector through its transversality and its linearity with respec to $Q$.

This is however not the only signature of a direct measurement of electromagnetic memory. In a realistic experiment, the emission of a charge (or scattering) takes place over a characteristic time scale that we will denote as $\Delta u$, at which the velocities of outgoing (and incoming) charges depart from their asymptotic constant value. Such a time scale is equivalently associated to a ``size'' of the emitted charges and effectively smoothens out the delta function in asymptotic retarded time within Eq.~\eqref{eq:ChargeCurrent}. This in turn results in a characteristic rise-time-scale of memory production by altering the Heaviside stepfunciton within the solutions in Eqs.~\eqref{eq:SolMemI} and \eqref{eq:SolMemII}. The corresponding transverse electric field in the radiation zone in Eq.~\eqref{eq:ElectricFieldtoA} will consequently be of a typical dome shape with obvious non-zero time integral, representing another characteristic signature of electromagnetic memory to look for in realistic observations \cite{Bieri:2023btq}.

Different experiment types have already been proposed for a direct measurement of electromagnetic memory, both in macroscopic and microscopic setups of memory production. One concrete proposal \cite{Bieri:2023btq} suggests the use of a dipole antenna generating two opposite unbound charge pulses that induce an emission of electromagnetic memory \footnote{Note that for a simultaneous emission of multiple current pulses in different directions, the results in section~\ref{sec:MemoryCalc} can simply be stacked. Therefore, an emission of opposite currents in a dipole antenna, for instance, would result in a memory production, in which the features of Fig.~\ref{fig:PlotEMem} are mirrored at $\theta=\pi/2$ with opposite signs.}. This signal could then be observed as a net current within a second receiver antenna, as an indication for a non-zero time integral of the radiative electric field. Another thought experiment \cite{Susskind:2015hpa} considers an ``explosion'' of charged-particles from a localized source surrounded by a sphere of superconducting nodes, within which the memory effect could be observed through a permanent imprint within the relative phases of the nodes.

With this work, we show that in similar experimental setups, the amplitude of electromagnetic memory can significantly be enhanced within a Lorentz breaking phase, whenever the superluminal regime $v_q\geq v_\text{em}$ is reached, forming a circle of amplitude enhancement about the direction of charge propagation at the characteristic angle
\begin{equation}
    \theta=\arccos(v_\text{em}/v_q)\,.
\end{equation}
As already mentioned, unlike in the gravitational analog, the structure of electromagnetic memory is such that an important increase in amplitude can already be achieved in a Lorentz preserving situation as proposed in \cite{Bieri:2023btq,Susskind:2015hpa}, when the speed of charge emission approaches the vacuum speed of light $c$. However, both the speed of a current pulse in a dipole antenna, as well as the emission velocity of charged particles for instance in $\beta$-decay \footnote{The idealized situation of a charge emitted at constant velocity $v_q$ considered in this work is indeed an excellent model of $\beta$-decay \cite{PhysRev.76.365,Jackson:1998nia}.} are not expected to reach the threshold of enhancement of $v_q>0.99 \,c$ identified in section~\ref{sSec:DirecitonsOfMemoryEnhancement}. In this context, a down scale of the effective phase velocity of light through Lorentz violating effects is crucial to reach the regime of significant memory amplification. 

To get a feeling for the significance of the enhancement, consider the values $|\Delta A^T|\sim \frac{q}{\text{[C]}}\frac{\text{[m]}}{r} 10^2 - 10^4$[Vs/m] in SI units presented in Fig.~\ref{fig:PlotEMem}, as a function of charge and distance. Both for the dipole antenna proposal in \cite{Bieri:2023btq} and typical $1$ MeV electron beams, a somewhat conservative value of the peak current is of the order of milliampere with the velocity scale close to $c$. In this case, and at typical laboratory length scales of meters, hence, timescales of nanoseconds, the charge of the currents are of order $q\sim 10^{-12}$[C]. With these numbers, a two-orders-of-magnitude gain in the potential amplitude lifts the expected strength of the electric field in Eq.~\eqref{eq:ElectricFieldtoA} from $E^\text{sub}_\text{max}\sim 10^{-3}\times10^2\,[\text{V/m}]\sim 10^{-1} $ [V/m] to $E^\text{sup}_\text{max}\sim 10$ [V/m], while the velocity kick on a single electron increases from $|\Delta v_\text{sub}|\sim 10 $ [m/s] to $|\Delta v_\text{sup}|\sim 10^3$ [m/s]. Note that at a given length scale, both the velocity kick, as well as the maximal electric field value, primarily depend on the scale of the emitted current, whose maximization should therefore be prioritized.

In general, a decrease of the phase velocity of light can either be realized within particular media, as explicitly presented in section~\ref{sec:MemoryCalc} in the case of a simple dielectric medium, or in the vacuum within adequate slow-wave structures, such as traveling-wave tubes (TWT). Of course, the conception of a realistic experimental setup to measure electromagnetic memory through the Lorentz breaking enhancement identified in this work still requires at least two significant steps that we want to address in future work. The first is a refined, quantitative computation of the memory amplitude within the given situation, taking into account finite-size- as well as expected shock-wave-effects that cannot be captured through the Green's function of the wave equation \footnote{The computational techniques developed in the context of Cherenkov radiation might be of great use.}. The second is an 
elaboration on the measurement technique, where it must in particular be ensured that the measurement time-scale is quick enough to capture the memory signal within the unbounded state of the emitted charges and that the memory kick can be discriminated from possible Poynting vector kicks associated with radiation pressure.

\section{Conclusion}\label{sec:Conclusion}

The computations in this work show that the amplitude of electromagnetic memory is significantly magnified, as soon as its past light cone, characterized by the phase velocity of light, is asymptotically approached by the spacetime direction of emission of a memory-sourcing charge. This condition can be met either:
\begin{enumerate}
    \item Around the direction of charge emission $\hat{\mathbf{n}}=\hat{\mathbf n}_q$ as the radial speed of emission asymptotes towards the speed of light $v_q\rightarrow v_\text{em}$. 
    \item Around the direction $\hat{\mathbf{n}}\cdot\hat{\mathbf n}_q=v_\text{em}/v_q$ for a superluminal emission $v_q> v_\text{em}$.
\end{enumerate}
While the latter case necessarily requires a reduction of the effective phase velocity of light $v_\text{em}<c$, the former can also be achieved at a vacuum speed of light $v_\text{em}=c$ with the emission of ultrarelativistic charges of $v_q\gtrsim 0.99 c$. 


These conclusions might open the door for the conception and implementation of realistic laboratory experiments to measure for the first time the phenomenon of electromagnetic memory.
Like in the gravitational case, a direct measurement of the phenomenon of memory would have important theoretical implications. The case of electromagnetic memory, however, comes with the advantage, that one could have a direct access to the true effect at vanishing frequency characterized by the net velocity kick. This would enable a direct observational verification of soft theorems and the concept of asymptotic symmetries \cite{He:2014cra,Pasterski:2015zua}. In contrast, in the gravitational case, such a direct link between the corners of the infrared triangle will be obscured by the fundamental limitation in frequency bands of the current and planned gravitational wave detectors.

Although the actual divergence in the memory amplitude as well as the emitted radiative energy remains intangible with the treatment of this work, the enhancement in the vicinity of the critical directions is physical, as shown by the presence of the entirely well-defined effect in the subluminal vacuum case, and indicates the formation of a shock-wave. A crucial future task will be to compute the realistic peak in memory offset increase in specific local Lorentz breaking setups and conceptualize concrete measurement strategies of the electromagnetic memory amplitude by addressing the shock nature of the phenomenon, as well as associated finite-size and possible dispersion and absorption effects \footnote{Note, however, that the consideration of a finite acceleration time of the emitted charges has no influence on the soft limit of the radiation and only serves as a high-frequency cutoff as shown in Appendix~\ref{sSec:IndirectObservation}.}.  

On the theory side, the work opens a novel research direction within the context of the infrared triangle \cite{He:2014cra,Campiglia:2015qka,Pasterski:2015zua,Kapec:2015ena,Strominger:2017zoo}. It raises the question of whether the enhancement of electromagnetic memory can be formulated in terms of Lorentz-violating extensions of the leading Weinberg photon soft theorem \cite{Weinberg_PhysRev.140.B516}, with corresponding implications on the asymptotic $U(1)$ symmetry. Moreover, since the enhancement arises from the universal Green’s function factor of the wave equation, the same mechanism is equally expected to be present in hereditary tail effects linked to sub-leading soft photon factors \cite{Laddha:2018myi,Sahoo:2018lxl,Sahoo:2020ryf} and their well-known gravitational analogs \cite{Blanchet:1992br,Blanchet:1993ec,Laddha:2018vbn}\footnote{Concretely, as the tail effects are tightly connected to the solution of the wave equation as well \cite{Saha:2019tub,Sahoo:2020ryf}, the amplifying Green's function factor in Eq.~\eqref{eq: Greensfunction factor} will also appear in the tail contributions, as can for instance explicitly be seen in Eq.(12) of Ref.~\cite{Laddha:2018vbn}.}.


\begin{acknowledgments}
I am very grateful to Vitor Cardoso and David Garfinkle for valuable input and comments on a first version of the manuscript. I also want to thank an anonymous referee for constructive comments and in particular for pointing out the potential application of the work on the tail effects associated to the sub-leading soft theorem. JZ is supported by funding from the Swiss National Science Foundation
(Grant No. 222346). The Center of Gravity is a Center of Excellence funded by the Danish National Research Foundation under Grant No. 184.
\end{acknowledgments}

\appendix

\section{Full Expressions of Electromagnetic Memory}\label{App:FullMemorySolution}

In this appendix, we explicitly write out the solutions for the electromagnetic memory in Eqs.~\eqref{eq:SolMemI} and \eqref{eq:SolMemII} in terms of polarization content of the vector potential
$A^T_{\hat\theta}\equiv A_i \hat\theta_i$ and $A^T_{\hat\phi}\equiv A_i \hat\phi_i$.
The standard spacial transverse and associated longitudinal basis vectors read
\begin{subequations}\label{eq:Def Transverse Vectors u v}
\begin{align}
\hat{\boldsymbol{\theta}}(\Omega)&=(\cos\theta \cos\phi,\,\cos\theta \sin\phi,\,-\sin\theta )\,,\\
\hat{\boldsymbol{\phi}}(\Omega)&=(-\sin\phi,\,\cos\phi,\,0)\,,\\
 \hat{\mathbf{n}}(\Omega)&=(\sin\theta \cos\phi,\,\sin\theta \sin\phi,\,\cos\theta )\,,\label{eq:Def Direction of Propagation n}
\end{align}
\end{subequations}
satisfying $\delta_{ij}=n_in_j+u_iu_j+v_iv_j$.

The memory component within the asymptotic radiation is linearly polarized, such that the residual polarization angle freedom $\psi$ in rotating about a given axis of propagation can be used to set the $\hat\phi$ polarization to zero. Concretely, a rotation in the transverse plane results in
\begin{subequations}
\begin{align}
     \tilde A^T_{\hat\theta}(\psi)&=\cos\psi A^T_{\hat\theta} -\sin\psi \tilde A^T_{\hat\phi}\,,\\
\tilde A^T_{\hat\phi}(\psi)&=\sin\psi A^T_{\hat\theta} +\cos\psi \tilde A^T_{\hat\phi}\,.
\end{align}
\end{subequations}
The angle $\psi$ that fulfills the conventions above is
\begin{widetext}
\begin{subequations}
\begin{align}
  \psi_{\hat\theta}&=\arctan\left[\frac{\cos\theta\sin\theta_q\cos[\phi-\phi_q]-\sin\theta\cos\theta_q}{\Sigma(\Omega,\Omega_q)},\frac{\sin\theta_q\sin[\phi-\phi_q]}{\Sigma(\Omega,\Omega_q)}\right]\,,
 \label{eq:psi}\\
\Sigma(\Omega,\Omega_q)&\equiv\sqrt{(\sin\theta\cos\theta_q-\cos\theta\sin\theta_q\cos[\phi-\phi_q])^2+\sin^2\theta_q\sin^2[\phi-\phi_q]}\,.
\end{align}
\end{subequations}
\end{widetext}
The resulting general solutions read
\begin{subequations}\label{eq:SolMemapp}
\begin{align}
    _{\myst{(I)}}\tilde A^T_{\hat\theta}(\psi_{\hat\theta})&=\frac{q\,\mu v_q}{4\pi r} \Theta(u-u_q)\Gamma(\Omega,\Omega_q)\,,\\
    _{\myst{(I)}}\tilde A^T_{\hat\phi}(\psi_{\hat\theta})&=0\,,\\
    {}_{\myst{(II)}}\tilde A^T_{\hat\theta}(\psi_{\hat\theta})&=
\frac{q\,\mu v_q}{4\pi r}
[\Theta(u-u_q)-1]\Gamma(\Omega,\Omega_q)\,,\\
     _{\myst{(II)}}\tilde A^T_{\hat\phi}(\psi_{\hat\theta})&=0\,,
\end{align}
\end{subequations}
with
\begin{subequations}
\begin{align}
\Gamma\equiv\frac{\Sigma(\Omega,\Omega_q)}{1-\frac{v_q}{v_\text{em}}(\cos\theta\cos\theta_q+\sin\theta\sin\theta_q\cos[\phi-\phi_q])}\,.
\end{align}
\end{subequations}
The total electromagnetic memory offset in each of the cases C $=$ (I), (II) is then defined as
\begin{align}\label{eq:TotalMemOffset}
    \Delta A^T(r,\Omega,\Omega_q)&\equiv {}_{\myst{C}}\tilde A_{\hat\theta}^T(u\rightarrow\infty)- {}_{\myst{C}}\tilde A_{\hat\theta}^T(u\rightarrow-\infty)\,.
\end{align}

For a source charge emitted in the $z$-direction, hence $\theta_q=0$, this expression reduced to the familiar form
\begin{align}
\Gamma(\Omega,\{0,0\})&=\frac{\sin\theta}{1-\frac{v_q}{v_\text{em}}\cos\theta}\,.
\end{align}
In this case, the total electromagnetic memory offset within the vector potential [Eq.~\eqref{eq:TotalMemOffset}] reads
\begin{align}\label{Appeq:TotalMemOffsetSimp}
    \Delta A^T(r,\Omega,\{0,0\})
    &=
    \frac{q\mu v_q}{4\pi r}
    \frac{\sin\theta}
    {1-\frac{v_q}{v_\text{em}}\cos\theta}\,,
    \;\,
    \frac{v_q}{v_\text{em}}\cos\theta\neq1\,.
\end{align}

\section{The Energy Spectrum of Electromagnetic Memory}\label{sSec:IndirectObservation}

Although electromagnetic memory has not yet been observed directly, the energy spectrum of phenomena which are directly related to the process of memory production, is well-known and experimentally verified. In particular, the idealized model of a charge emitted at constant velocity $v_q$ from a localized source, discussed in section~\ref{sec:MemoryCalc}, has successfully been applied to beta decay \cite{PhysRev.76.365,Jackson:1998nia}. The reason for the accuracy of a classical treatment of such low-frequency effects is that the number of emitted photons diverges in the soft limit \cite{Weinberg_PhysRev.140.B516}. These experiments are characterized by the energy radiated per unit solid angle $d\Omega$ per unit frequency interval $d\omega$
\begin{align}\label{eq:EnergySpectrum}
    \frac{d^2 E}{d\omega d\Omega} = \frac{2 r^2}{v_\text{em}\mu}|\tilde{\mathbf{E}}(\omega)|^2\,,
\end{align} 
where a tilde denotes the Fourier transform.

The radiative electric field associated with the Lorentz breaking memory solutions in Eqs.~\eqref{eq:SolMemI} and \eqref{eq:SolMemII} can for $\hat{\mathbf n}\cdot\hat{\mathbf n}_q
    \neq v_\text{em}/v_q$ be written as
\begin{align}\label{eq:TotalE}
    E^i_T(u,r,\Omega)
    &=
    -\frac{q\mu v_q}{4\pi r}
    \frac{
    D(u-u_q)\,\perp^{ij}\hat n_q^j
    }{
    1-\frac{v_q}{v_\text{em}}
    \hat{\mathbf n}\cdot\hat{\mathbf n}_q
    }\,,
\end{align}
where to be concrete, we will associate the high-frequency cutoff $\Delta u$ in the memory production to the width of a Gaussian function
\begin{equation}
    D(u-u_q)\equiv \frac{1}{\Delta u\sqrt{2\pi}}e^{-\frac{(u-u_q)^2}{2\Delta u^2}}\,.
\end{equation}
The energy spectrum in Eq.~\eqref{eq:EnergySpectrum} then becomes
\begin{align}
    \frac{d^2 E}{d\omega d\Omega} = \frac{\mu q^2}{8 \pi^2 }\frac{v_q^2}{v_\text{em}}\frac{|\tilde D(\omega)|^2\,(1-(\hat{\mathbf{n}}\cdot\hat{\mathbf n}_q)^2)}{(1-\frac{v_q}{v_\text{em}}\hat{\mathbf{n}}\cdot\hat{\mathbf n}_q)^2}\,,\label{eq:IntensityPer}
\end{align} where explicitly, the Fourier transform results in
\begin{equation}
    \tilde{D}(\omega)=\int_{-\infty}^\infty \frac{d u}{\sqrt{2\pi}}\, D(u-u_q)e^{-i\omega u}=\frac{e^{-\frac{\Delta u^2\omega^2}{2}+iu_q\omega}}{\sqrt{2\pi}}\,.
\end{equation}
In the soft limit $\omega\rightarrow 0$ and choosing $\theta_q=0$, the result is directly proportional to the total memory offset in Eq.~\eqref{eq:TotalMemOffsetSimp}
\begin{align}\label{eq:EnergyPerSoft}
    \frac{d^2E}{d\omega d\Omega}\Bigg|_{\omega\rightarrow 0} = \frac{ r^2}{v_\text{em}\mu\pi} \left[\Delta A^T\right]^2\,.
\end{align}

Moreover, for $v_q<v_\text{em}$, the expression in Eq.~\eqref{eq:IntensityPer} can be integrated over the angles and the frequencies to obtain 
\begin{equation}\label{eq:TotEMem}
    E_\text{tot}\simeq \frac{\mu q^2}{8\pi^{3/2}} \frac{\gamma_q m_q c^2}{ \hbar}v_\text{em}\left(\frac{v_\text{em}}{v_q}\ln\left[\frac{1+v_q/v_\text{em}}{1-v_q/v_\text{em}}\right]-2\right)\,.
\end{equation}
In this estimate of the energy emitted in electromagnetic memory radiation, we have used that for the emission of a single charged particle of mass $m_q$, the emission time scale $\Delta u$ can be estimated through the Heisenberg uncertainty principle to be of order $\Delta u\sim \frac{\hbar}{E_q} =\frac{\hbar}{\gamma_q m_q c^2}$.
As a conclusion, the magnification in the memory amplitude, directly translates into an enhancement of the total emitted energy, with the biggest support coming from the low frequency spectrum and the total radiative energy diverges as $v_q\rightarrow v_\text{em}$. 

The expressions in Eqs.~\eqref{eq:EnergyPerSoft} and \eqref{eq:TotEMem} recover known results in the Lorentz preserving case $v_\text{em}=c$ \cite{PhysRev.76.365,Jackson:1998nia,PhysRevD.68.084011}, upon converting to Gaussian units in which $\mu_0=4\pi/c^2$. The fact that these energy spectra of electromagnetic radiation associated to $\beta$-decay are observationally confirmed (see \cite{Jackson:1998nia,Italiano_2024} and references therein) provides additional support for the physicality of the memory enhancement along the critical spacetime directions identified in this work. Moreover, the divergence present in the memory offset as $v_q\geq v_\text{em}$ is a true divergence that equally appears in the amount of energy emission, calling for a resolution in finite-size effects.

\section{Emission Along Lightcone Directions}\label{App:LightconeEmission}

We explicitly show that the condition of large memory production in Eq.~\eqref{eq:LightconeCond} can geometrically be understood as the directions in which the asymptotically free charge is emitted along a past null cone of memory evaluation. For this purpose, we will compute the intersection between the future causal cone of the emission of a charge with velocity $v_q$ at the origin
\begin{equation}
    x^2+y^2+z^2-v_q^2 t^2=0\,,
\end{equation}
and the past light-cone of a memory event at some given distance $r v_{\text{em}}/c$ characterized by the speed $v_\text{em}$
\begin{equation}
    (x-r v_{\text{em}}/c)^2+y^2+z^2-(v_\text{em}t-rv_{\text{em}}/c)^2=0\,.
\end{equation}

Note that for concreteness, we have choose here to evaluate the memory in the $x$ direction, hence 
\begin{equation}\label{eq:directionx}
    \hat{\mathbf{n}}=(1,0,0)\,.
\end{equation}
In contrast to the discussion in section~\ref{sSec:DirecitonsOfMemoryEnhancement} we therefore fix the direction of memory evaluation and vary the radial vector of charge emission $\hat{\mathbf n}_q$.
Moreover, for simplicity, we will set the irrelevant direction $y$ to zero.
The intersection line between the cones is then computed to be given by the solutions
\begin{subequations}
\begin{align}
    x_c&=tv_\text{em}+\frac{t^2(v_q^2-v_\text{em}^2)}{2 r v_\text{em}/c}\,,\\
    z_c&=\pm\sqrt{t^2 v_q^2-\left\{t v_\text{em}+\frac{t^2(v_q-v_\text{em})(v_q+v_\text{em})}{2 r v_\text{em}/c}\right\}^2}\,.
\end{align}
\end{subequations}

\begin{figure}[h]
  \centering
    \includegraphics[scale=0.58]{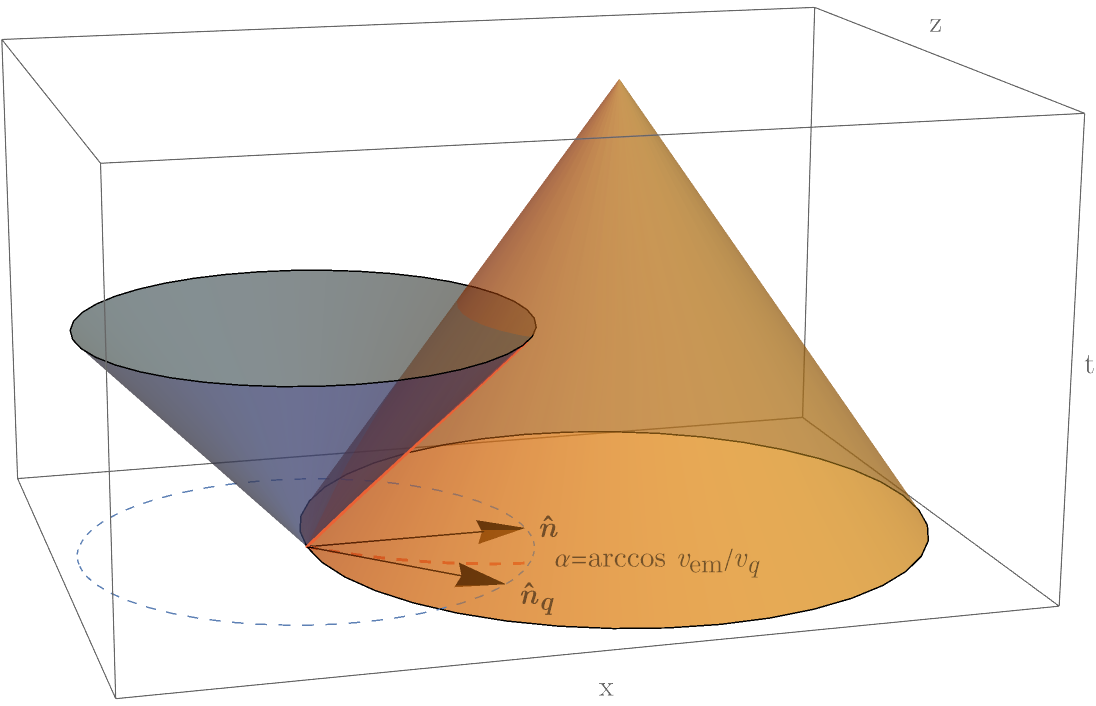}
  \caption{\small Causal cone of emission of a charge $q$ at a speed $1>v_q>v_\text{em}$ with $c=1$ (blue) intersecting a past null cone of a memory event (orange) admitting critical directions of memory enhancement. The angle $\alpha$ precisely corresponds to the critical angle identified in Eq.~\eqref{eq:LightconeCond}.}\label{fig:LCones}
\end{figure}

As expected, there is only a real solution if $v_q\geq v_\text{em}$. This can be seen more clearly by explicitly computing the normalized spacial projection of the emission direction of the charge $\hat{\mathbf{n}}_q^c=(v_\text{em}/v_q,0,\sqrt{1-v_\text{em}^2/v_q^2})$ that corresponds to an emission along the lightcone directions $\partial_t x_c|_{t=0}$ and $\partial_t z_c|_{t=0}$.
The critical direction of charge emission $\hat{\mathbf{n}}^c_q$ along the intersection of the past causal light cone defined through a direction of memory evaluation $\hat{\mathbf{n}}$ in Eq.~\eqref{eq:directionx} is therefore characterized by a scalar product satisfying $\hat{\mathbf{n}}\cdot\hat{\mathbf{n}}^c_q=\frac{v_\text{em}}{v_q}$.
This condition precisely coincides with the critical directions of diverging memory amplitude in Eq.~\eqref{eq:LightconeCond} identified in section~\ref{sSec:DirecitonsOfMemoryEnhancement}.
The corresponding geometrical configuration is depicted in Fig.~\ref{fig:LCones}.

\bibliographystyle{utcaps}
\bibliography{references}

\end{document}